\documentclass[twocolumn, trackchanges]{aastex62}
\usepackage{booktabs}
\usepackage{amsmath,amstext} 

\begin{document}

%\title{Investigating Outflow Structures of the Binary Neutron Star Merger GW170817 and Cosmological Short Gamma-ray Bursts}
%\title{GW170817 Afterglow Reveals that all Short Gamma-Ray Bursts are Neutron Star Mergers}
\title{GW170817 Afterglow Reveals that Short Gamma-Ray Bursts are Neutron Star Mergers}
% \title{BNS GW170817 is same as cosmological short GRBs}
% \title{BNS GW170817 is a cosmological short GRB viewed off-axis}
% \title{Cosmological short GRB are like BNS GW170817 viewed on-axis}
% \title{Cosmological short GRB are like GW170817 viewed on-axis}
% \title{BNS GW170817 is like cosmological short GRBs for an on-axis observer}
% \title{BNS GW170817 shares outflow structures with short GRBs}

\author{Yiyang Wu}
\affil{Center for Cosmology and Particle Physics, New York University}

\author{Andrew MacFadyen}
\affil{Center for Cosmology and Particle Physics, New York University}

\begin{abstract}
We systematically investigate the outflow structure of GW170817 in comparison with a sample of 27 cosmological short GRBs by modelling their afterglow light curves. 
We find that cosmological short GRBs share the same outflow structures with GW170817, relativistic structured jets. The jet opening angle of GW170817 is $6.3^{+1.1}_{-0.6}{}^{\circ}$, which is consistent with that of cosmological short GRBs ($\theta_0 = 6.9^\circ \pm 2.3^{\circ}$). Our analysis indicates that GW170817 is viewed off-axis ($\theta_{\rm obs} = 30^{+7}_{-4}{}^{\circ}$), while cosmological short GRBs are viewed on-axis ($\theta_{\rm obs}  \lesssim \theta_0$). The exceptional properties of the GW170817 afterglow can be explained by the difference in observation angle alone. 
We demonstrate that the light curves of the GW170817 afterglow, if viewed on-axis, are consistent with those of cosmological short GRBs. Other properties of GW170817, such as Lorentz factor $\Gamma \approx 150$, spectral index $p \approx 2.15$, isotropic equivalent energy $E_{\rm iso} \approx 8 \times 10^{52}$ erg and interstellar medium density $n_0 \approx 10^{-2}$ proton cm$^{-3}$, fit well within the ranges of those of cosmological short GRBs. 
The similarity between the  GW170817 outflow structure and those of cosmological short GRBs indicates that cosmological short GRBs are likely neutron star mergers. 

\end{abstract}


\section{Introduction}
On 17 August 2017, LIGO/Virgo detected the first binary neutron star (BNS) merger event, known as GW170817 \citep{Abbott2017}. Approximately 1.7 seconds later, the \emph{Fermi} space telescope detected a weak short-duration gamma-ray burst (GRB), GRB170817A, with an inferred sky location coinciding with that of GW170817 \citep{Goldstein2017, Savchenko2017}. After intensive multiband monitoring, a long-lived GRB afterglow was detected at radio, optical and X-ray wavelengths \citep{Alexander2017, Haggard2017, Hallinan2017, Kasliwal2017, Margutti2017, Troja2017, Alexander2018, Dobie2018, Lyman2018, Margutti2018, mooley2018mildly, mooley2018strong, nynka2018fading, piro2018long, Resmi2018, ruan2018brightening, van2018year,lamb2019optical}. 
%The temporal and spatial connections between GW170817 and GRB170817A provide convincing evidence that BNS mergers are associated with short GRBs.
 
Compared to classical short GRBs, the $\gamma$-ray emission and the afterglow from GW170817 displayed exceptional properties. Located in NGC 4993, an elliptical galaxy at a distance of $39.5$ Mpc ($z = 0.00973$), it is the closest burst among short GRBs with host galaxy identifications and has the lowest total gamma-ray energy $\sim 10^{46} \text{-} 10^{47}$ erg \citep{fong2017electromagnetic, Goldstein2017, Savchenko2017}. For comparison, classical short GRBs are at cosmological distance and typically have $\gamma$-ray energies of  $\sim 10^{50} \text{-} 10^{52}$ erg  \citep{fong2015decade}. The afterglow from GW170817 had a late onset at $\sim$ 9 days \citep{Margutti2017, Troja2017} and a steady brightening up to  $\sim 100$ days \citep{Hallinan2017, Lyman2018, mooley2018mildly, ruan2018brightening}. The afterglows from classical short GRBs are typically detected shortly after prompt emission and display a general decline (sometimes accompanied with short-lived plateaus and flares)  \citep{fong2015decade}. 

%Two leading theories were proposed to explain these exceptional behaviors. The first one is that GW170817 is a new astrophysical transient with intrinsically low luminosity. It has a mildly relativistic quasi-spherical outflow and is powered by the interaction between the choked jet and the BNS cloud. This scenario was extensively discussed by \citet{Bromberg2017, Kasliwal2017, Hotokezaka2018,  Gottlieb2018, mooley2018mildly, nakar2018gamma, Xie2018}. The other theory is that GW170817 is viewed off-axis and would be seen as a classical short GRB of typical luminosity for an on-axis observer. The resulting outflow is a relativistic structured jet viewed off-axis \citep{Kathirgamaraju2017, Lamb2017, lazzati2017, beniamini2018lesson, d2018evidence, Gill2018,  Lyman2018, Margutti2018, Resmi2018, Troja2018, Xie2018}. A key question has raised whether GW170817 has a mildly relativistic quasi-spherical outflow resulted from a new class of astrophysical transient or a relativistic structured jet resulted from a classical short GRB. 

Two leading models were proposed to explain these exceptional behaviors of GW170817: a relativistic structured jet viewed off-axis \citep{Kathirgamaraju2017, Lamb2017, Alexander2018, beniamini2018lesson, d2018evidence, Gill2018, lazzati2017, Lyman2018, Margutti2018, Resmi2018, Troja2018, Xie2018} and a mildly relativistic quasi-spherical outflow \citep{Bromberg2017, Kasliwal2017, Hotokezaka2018,  Gottlieb2018, mooley2018mildly, nakar2018gamma, Xie2018}. A heated debate concerning the post-merger outflow structure was raised, since these two models, though significantly different, both succeeded in explaining the observed late onset and early brightening.  
 %These two models, though significantly different, both succeeded in explaining the observations. A heated debate concerning the post-merger outflow structure was raised. 
 
\citet{wu2018constraining} analyzed the multiband GW170817 afterglow data with the physically motivated analytic two-parameter  ``boosted fireball'' model for the outflow structure after it has expanded many orders of magnitude larger than the scale of the central engine \citep{Duffell2013RT}. This model encompasses a family of outflows with structures varying smoothly from a highly collimated ultra-relativistic jet to an isotropic outflow. %By performing MCMC analysis in an eight-dimensional parameter space, different kinds of outflow structures could be explored and thus the two leading outflow structures could be naturally distinguished. The fitting results favored the relativistic structured jet viewed off-axis rather than the quasi-spherical outflow. 
 By performing MCMC analysis, these two leading outflow structures, along with general outflow structures, can be directly compared and distinguished. The fitting results favored the relativistic structured jet viewed off-axis and the quasi-spherical outflow was ruled out due to significantly larger reduced $\chi^2$.
 
Several other studies also supported the relativistic structured jet model. \citet{lamb2018late} demonstrated that two models have different behaviors with respect to the decline of the post-peak afterglow. The observed steep decline indicates the relativistic structured jet \citep{lamb2018late, van2018year,lamb2019optical}. \cite{mooley2018strong} reported Very Long Baseline Interferometry (VLBI) observations, which  indicate a superluminal proper motion of the radio counterpart of GW170817. \cite{2018ApJ...865L...2Z} analyzed the proper motions for the two leading outflow structures and found that both outflows are consistent with the VLBI observations. 

%As a local short GRB, GW170817 only serves as one burst with a jet-like structure. 
Given these extensive studies, it is generally accepted that GW170817 has a relativistic jet-like structure, which leads us to ask: if GW170817 is a typical short GRB viewed off-axis, do GW170817 and short GRBs, in general, share similar outflow structures? Are all cosmological short GRBs are neutron star mergers? 

In this letter, we present a comprehensive comparison between GW170817 and the short GRB population. We apply the same tools developed in \citet{wu2018constraining} and directly compare the outflow structures of GW170817 with those of a sample population of short GRBs \citep{fong2015decade}. In Section \ref{sec: GRB method}, we give a brief overview of the boosted fireball model. Section \ref{sec: GRB data} describes the dataset of 27 cosmological short GRBs. The results are summarized in Section \ref{sec: GRB results} and discussed in Section \ref{sec: GRB discussion}.

\section{Method}\label{sec: GRB method}
The idea of boosted fireball model is that a fireball of specific internal energy $\eta_0$ is launched with a boost Lorentz factor $\gamma_B$ (for details see \citet{Duffell2013, wu2018constraining}). Due to relativistic beaming, the outflow has a characteristic Lorentz factor $\Gamma \sim 2\eta_0\gamma_B$ and a characteristic jet opening angle $\theta_0 \sim 1/\gamma_B$. Depending on the two parameters, $\eta_0$ and $\gamma_B$, a family of outflow structures can be generated, from a highly collimated ultra-relativistic jet to an isotropic fireball. Because of its flexibility, the boosted fireball model can serve as a generic outflow model, which is suitable for investigating a population of cosmological short GRBs. 

The parameter space of the boosted fireball model consists of hydrodynamic parameters ($\eta_0$, $\gamma_B$, the explosion energy $E_{0}$, and the interstellar medium (ISM) density $n_0$), radiation parameters (the spectral index $p$, the electron energy fraction $\epsilon_e$ and the magnetic energy fraction $\epsilon_B$), and observational parameters (the observation angle $\theta_{\rm obs}$). By performing an MCMC analysis in this parameter space, we can explore a family of outflows viewed from different observation angles and automatically find the best-fitting parameters.  

To enhance fitting performance, $E_0$ and $n_0$ are made dimensionless, $E_{0,50}\equiv E_0/10^{50}\text{ erg}$ and $n_{0,0}\equiv n_0/1\text{ proton cm}^{-3}$, and are transformed into a logarithmic scale. The boundaries of the parameter space are $\log_{10}E_{0,50} = $ [-6,3], $\log_{10}n_{0,0} = $ [-6, -1], $\eta_0 = $ [2, 10], $\gamma_B = $ [1, 12], $\theta_{\rm obs} = $ [0, 1], $\log_{10}\epsilon_e = $ [-6, 0], $\log_{10}\epsilon_B = $ [-6, 0] and $p = $ [2, 3]. Since most short GRBs occur in a low-density environment \citep{fong2015decade}, we set the boundaries of density to be $\log_{10}n_{0,0} = $  [-6, -1]. The upper boundaries of $\eta_0$ and $\gamma_B$ are limited by the expense of the hydrodynamics simulations. Higher Lorentz factors are computationally expensive for parameter space study. Considering quasi-spherical outflows usually have wide opening angles corresponding to  $\gamma_B \sim 1\text{-}2$, our parameter space is large enough to distinguish jet-like and quasi-spherical structures. 

% \begin{table}[hbt]\label{tab2: parameter}
% \centering
% \caption{Boundaries of the Parameter Space}
% \begin{tabular}{@{}cc@{}}
% \toprule
% Parameter     & Range                           \\             \midrule
% $\log_{10}E_{0,50}$  &  [-6, 3] \\
% $\log_{10}n_{0,0}$       & [-6, -1]   \\
% $\eta_0$      &  [2, 10]                     \\
% $\gamma_B$    & [1, 12]                               \\ 
% $\theta_{\rm obs}$ & [0, 1]                         \\
% $\log_{10}\epsilon_e$ & [-6, 0]  \\
% $\log_{10}\epsilon_B$ & [-6, 0]  \\
% $p$ &  [2, 3]  \\  \bottomrule
% \end{tabular}
% \end{table}

By making use of the scaling relations in the hydrodynamic and radiation equations \citep{Eertenn2012Scale, Ryan2015}, we are able to generate synthetic light curves in milliseconds, which allows us to perform MCMC fitting in a reasonable amount of time. However, scaling relations also result in degeneracies between $E_0$, $n_0$, $\epsilon_e$ and $\epsilon_B$. In practice, we observe broad posterior distributions for these parameters. Even though degenerate parameters exist in our analysis, other parameters ($\eta_0$, $\gamma_B$, $\theta_{\rm obs}$ and $p$) are robustly constrained. The uncertainties of degenerate parameters can be incorporated into the marginalized distributions of non-degenerate parameters. In \cite{wu2018constraining}, we demonstrated that the medians of marginalized distributions under two scenarios, free and fixed density, were consistent. 

Samples are generated by the parallel-tempered affine-invariant ensemble sampler implemented in the \texttt{emcee} package \citep{Goodman2010,Foreman2013}. We set $10$ temperature levels and $100$ walkers per level for the sampler. The walkers are initialized in a small ball near the maximum of the posterior, calculated through trial runs. We drop the first 5,000 steps as burn-in and perform analysis on the following 5,000 steps.

\section{Data} \label{sec: GRB data}
We consider a catalog of afterglow observations, consisting of all short GRBs from 2004 November to 2015 March with prompt follow-up observations \citep{fong2015decade}. Redshifts of these bursts span from $z=0.12$ to $z=2.6$. The observational data of GW170817 is taken from \citet{ Alexander2018,Margutti2018, van2018year}. 

Afterglow light curves of short GRBs are sometimes subject to early-time effects, such as steepenings (GRBs 051221A and 111020A), plateaus (GRB 051221A) and flares (GRBs 050724A and 111121A). Since the boosted fireball model assumes the outflow has already expanded far from the central engine, it is not designed to explain these early-time features, which could contaminate the afterglow emission and significantly affect the fit. Thus, we trim the early light curves to ensure the model is applied to the appropriate regime. 

Due to the small number of well observed short GRB afterglows, we would like to include as many short GRBs as possible. Even though fits are performed in an eight-dimensional parameter space, we restrict our analysis to all known 27 short GRBs with at least 6 data points. For 13 short GRBs that do not have a determined spectroscopic redshift, we assume $z=0.46$, set by the median of the short GRBs with known redshifts \citep{fong2017electromagnetic}. Of the 27 short GRBs, there are 26 X-ray detections, 23 optical/near-infrared detections and 4 radio detections. Four bursts have detections in all three bands. Eighteen bursts have both X-ray and optical/near-infrared detections. Five bursts are detected in only one band.  

\section{Results} \label{sec: GRB results}
\subsection{Goodness of Fit}
We perform MCMC analysis on the afterglow light curves of GW170817 and 27 short GRBs. 
The quality of the fits varied from burst to burst. For light curves with enough data points, we can use $\chi^2$/DOF to determine the goodness of fit.  To be counted as a good
fit, we require $\chi^2$/DOF $\leq$ 3 for bursts with enough data points.  Since we allow the number of data points to be less than the number of dimensions of the parameter space in order to incorporate more bursts, the degrees of freedom can be zero or even negative, which makes $\chi^2$/DOF meaningless. Thus, we use $\chi^2$ to determine the goodness of fit and require $\chi^2 \leq 10$ for a good fit. However, there are cases with low  $\chi^2$ or $\chi^2$/DOF, but the fitting light curves are choppy and subject to overfitting, which we consider bad fits.
%In Table \ref{tab: goodness}, we show the goodness of fit ($\chi^2$,  $\chi^2$/DOF) for GW170817 and 27 bursts.
%In some cases, $\chi^2$ is pretty large, but fitted light curves smoothly match most data points, which we consider good fits. In other cases,  $\chi^2$ is small, but the fitted light curves are choppy and subject to overfitting, which we consider bad fits.  
 
 We find GW170817 and 14 bursts have reasonably good fits. For 13 bursts, we are unable to find good fits. There are several factors that could lead to a low quality fit: a lack of enough data points, too much noise in the data, the quality of synthetic light curves and violations of model assumptions, such as homogeneous ISM. %In the Appendix \ref{tab: goodness}, we show the goodness of fit ($\chi^2$,  $\chi^2$/DOF) for GW170817 and 27 bursts.

% In Figure \ref{fig: OneGood}, we show the MCMC results for GRB 060313. It serves as an example of a particular good fit. The diagonal plots show the one-dimensional marginalized posterior distribution for each parameter and the off-diagonal plots show two-dimensional posterior surface for each pair of parameters. We find tight constraints for fitting parameters: $\eta_0 = 9.60^{+0.32}_{-0.90}$, $\gamma_B = 9.54^{0.50}_{-0.67}$ and $\theta_{\rm obs} = 0.01^{+0.01}_{-0.00}$. The corresponding characteristic parameters are $\theta_0 = 0.105^{+0.005}_{-0.008}$ and $\Gamma = 183^{+15}_{-30}$. Thus, this burst has a relativistic jet-like outflow structure and is viewed on-axis. The inset shows the best-fitting light curves in both X-ray and optical bands. The reduced $\chi^2$ is 0.81 (28.3/35). 

% \begin{figure*}[hbt] 
% \centering
% \includegraphics[width=\linewidth]{R_060313_1_ContourLC.pdf}
% \caption{MCMC analysis results for GRB060313. The corner plots show the one-dimensional (diagonal) and two-dimensional (off-diagonal) projections of the posterior distributions for fitting parameters. The dashed lines indicate the medians and the symmetric 68\% uncertainties. Inset shows the best-fitting light curves (solid lines) and observational data (circles with error bars).}
% \label{fig: OneGood}
% \end{figure*}

\subsection{Constraints on Fitting Parameters}
\begin{figure}[hbt] 
\centering
\includegraphics[width=\linewidth]{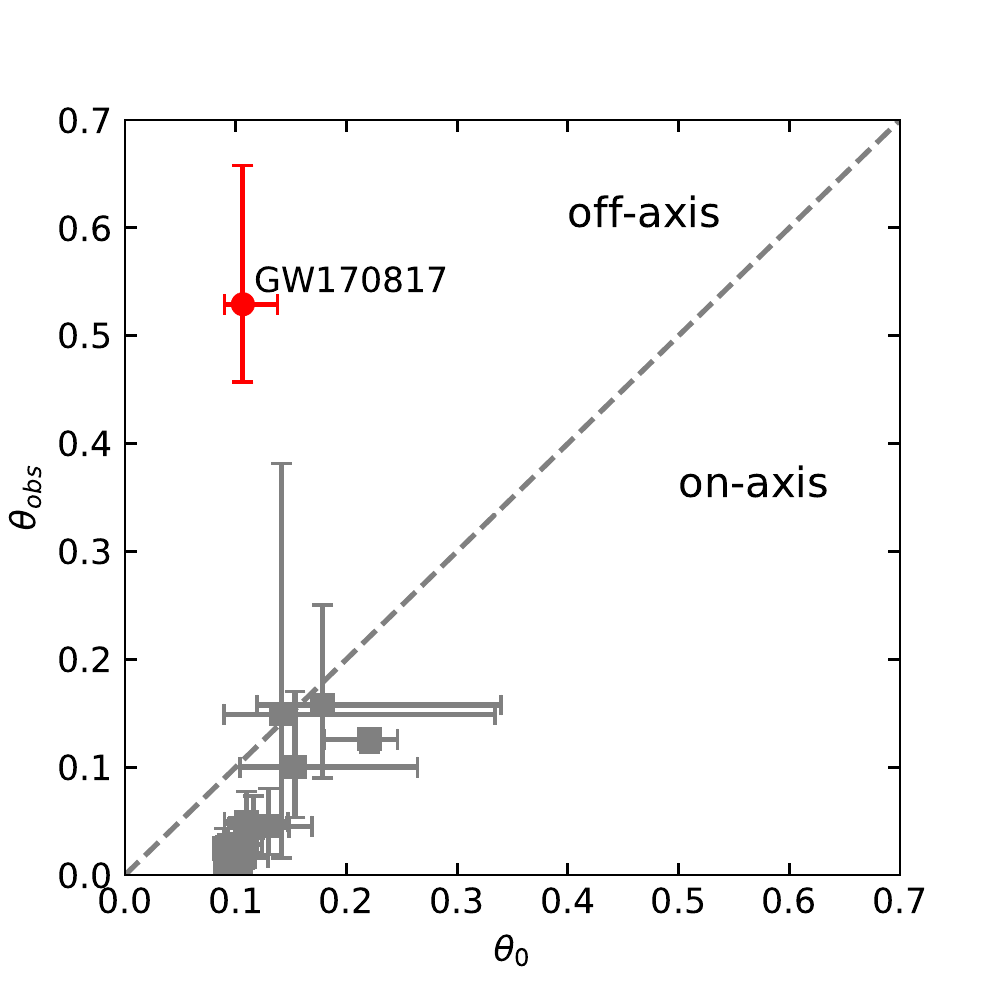}
\caption{Fit results of jet opening angle $\theta_0$ and observation angle $\theta_{\rm obs}$ for GW170817 (red circle) and 14 short GRBs (blue squares) in the plane ($\theta_0, \theta_{\rm obs}$). Jet opening angles are estimated as $\theta_0 \sim 1/\gamma_B$.  Markers with error bars indicate median values and symmetric 68\% quantiles. The grey dashed line indicates $\theta_{\rm obs} = \theta_0$. }
\label{fig:contour_theta}
\end{figure}

 In Table \ref{tab: results}, we show the constraints of fitting parameters ($\eta_0, \gamma_B, \theta_{\rm obs}, p, E_{0,50}, n_{0,0}, \epsilon_e, \epsilon_B$) and corresponding characteristic parameters ($\theta_0, \Gamma$) for GW170817 and 14 good fit bursts.  In Figure \ref{fig:contour_theta}, we show the distribution of GW170817 (red circle) and 14 short GRBs (blue squares) in the ($\theta_0, \theta_{\rm obs}$)  plane.  GW170817 is found to have a jet opening angle $\theta_0 = 0.11^{+0.02}_{-0.01}$ radians $= 6.3^{+1.1}_{-0.6}{}^\circ$. Remarkably, all short GRBs are also found to have similar jet-like outflows. 
The mean and standard deviation of jet opening angles is 0.12 $\pm$ 0.04 radians = $6.9^\circ \pm 2.3^{\circ}$, which is consistent with that of GW170817.  
For bursts with distinct jet breaks, jet opening angles can be estimated from the observed jet breaks. 
\citet{fong2015decade} estimated the jet opening angle ($\theta_0 = 3^\circ \text{-} 8^\circ$) for GRB 111020A from its jet break, which is consistent with our value $\theta_0 = 4.7^\circ \text{-} 6.5^\circ$. 
$\theta_0$ may be lower since it is limited by the upper boundary of $\gamma_{B, \text{max}} = 12$, which corresponds to $\theta_{0, \text{min}} = 0.08$ radians $= 4.6^\circ$. GW170817 and short GRBs share the same outflow structures with $\gamma_B \gtrsim 9$ corresponding to structured jets, and are outside the range $\gamma_B =  1\text{-}2$ corresponding to quasi-spherical outflows.

  %we mistook the peaks of posterior distributions as medians.  In this letter, we correct the meid as 50\% quantiles. The values of fitting parameters could vary slightly depending on the shape of distributions. The main conclusion, a relativistic structured jet viewed off-axis, remains unaffected.
 \begin{table*}[hbt] 
\centering
\caption{Parameter Constraints for GW170817 and the 14 Cosmological Short GRBs}
\label{tab: results}
\begin{tabular}{ccccccccccccccc}
\toprule
{} &  $\eta_0$ &  $\gamma_B$ &  $\theta_{\rm obs}$ & $p$ &  $\theta_0$ & $\Gamma$ & $\log_{10} E_{0,50}$ & $\log_{10} n_{0,0}$ & $\log_{10} \epsilon_e$ & $\log_{10} \epsilon_B$ \\
\midrule
GW170817\footnote{We have corrected the medians in \citet{wu2018constraining}, which misreported the peaks of posterior distributions as the medians.} & $7.9^{+1.3}_{-1.4}$ &  $9.4^{+1.7}_{-2.1}$ &  $0.529^{+0.129}_{-0.072}$ &  $2.15^{+0.01}_{-0.01}$ &  $0.11^{+0.02}_{-0.03}$ &  $149^{+57}_{-54}$ &  $-0.2^{+0.8}_{-0.8}$ &  $-2.0^{+0.7}_{-1.0}$ &  $-1.0^{+0.6}_{-0.9}$ &  $-3.6^{+1.3}_{-1.4}$ \\ 
050709 & $9.4^{+0.4}_{-0.8}$ &  $4.5^{+1.0}_{-0.5}$ &  $0.126^{+0.008}_{-0.012}$ &  $2.87^{+0.01}_{-0.01}$ &  $0.22^{+0.04}_{-0.03}$ &  $85^{+23}_{-15}$ &  $-0.1^{+0.4}_{-0.6}$ &  $-1.6^{+0.4}_{-0.6}$ &  $-0.3^{+0.2}_{-0.3}$ &  $-3.4^{+0.9}_{-0.9}$ \\ 
050724A & $9.0^{+0.8}_{-2.4}$ &  $11.2^{+0.6}_{-1.3}$ &  $0.025^{+0.019}_{-0.014}$ &  $2.24^{+0.14}_{-0.18}$ &  $0.09^{+0.00}_{-0.01}$ &  $200^{+29}_{-70}$ &  $-1.1^{+0.1}_{-0.1}$ &  $-1.2^{+0.2}_{-0.3}$ &  $-0.4^{+0.1}_{-0.1}$ &  $-2.2^{+0.3}_{-0.3}$ \\ 
060313 & $9.6^{+0.3}_{-0.9}$ &  $9.5^{+0.5}_{-0.7}$ &  $0.007^{+0.006}_{-0.004}$ &  $2.09^{+0.04}_{-0.03}$ &  $0.10^{+0.01}_{-0.01}$ &  $183^{+16}_{-29}$ &  $-1.4^{+0.1}_{-0.1}$ &  $-1.1^{+0.1}_{-0.1}$ &  $-0.1^{+0.1}_{-0.1}$ &  $-1.7^{+0.1}_{-0.1}$ \\ 
061006 & $7.8^{+1.6}_{-2.1}$ &  $5.6^{+2.8}_{-2.7}$ &  $0.158^{+0.093}_{-0.068}$ &  $2.25^{+0.25}_{-0.16}$ &  $0.18^{+0.06}_{-0.16}$ &  $88^{+70}_{-54}$ &  $-0.6^{+0.7}_{-0.6}$ &  $-1.5^{+0.4}_{-0.6}$ &  $-0.6^{+0.4}_{-0.8}$ &  $-0.9^{+0.6}_{-0.6}$ \\ 
061201 & $6.3^{+2.6}_{-3.3}$ &  $7.1^{+4.1}_{-4.1}$ &  $0.149^{+0.232}_{-0.133}$ &  $2.55^{+0.27}_{-0.34}$ &  $0.14^{+0.05}_{-0.19}$ &  $90^{+111}_{-71}$ &  $0.5^{+1.6}_{-2.1}$ &  $-1.9^{+0.7}_{-1.4}$ &  $-0.8^{+0.6}_{-1.1}$ &  $-1.2^{+0.8}_{-1.9}$ \\ 
070724A & $7.3^{+1.9}_{-2.3}$ &  $6.5^{+3.1}_{-2.7}$ &  $0.100^{+0.070}_{-0.047}$ &  $2.39^{+0.30}_{-0.24}$ &  $0.15^{+0.05}_{-0.11}$ &  $95^{+81}_{-57}$ &  $-0.8^{+0.9}_{-0.6}$ &  $-1.4^{+0.3}_{-0.4}$ &  $-0.6^{+0.4}_{-0.8}$ &  $-0.5^{+0.4}_{-0.6}$ \\ 
070809 & $7.9^{+1.4}_{-2.1}$ &  $7.7^{+1.7}_{-1.8}$ &  $0.045^{+0.035}_{-0.026}$ &  $2.05^{+0.05}_{-0.03}$ &  $0.13^{+0.02}_{-0.04}$ &  $122^{+53}_{-53}$ &  $-1.4^{+0.5}_{-0.4}$ &  $-1.2^{+0.2}_{-0.4}$ &  $-0.2^{+0.2}_{-0.4}$ &  $-2.1^{+0.4}_{-0.6}$ \\ 
080426 & $9.9^{+0.1}_{-0.2}$ &  $11.1^{+0.6}_{-0.7}$ &  $0.004^{+0.003}_{-0.002}$ &  $2.24^{+0.11}_{-0.08}$ &  $0.09^{+0.00}_{-0.01}$ &  $219^{+15}_{-18}$ &  $-2.0^{+0.1}_{-0.1}$ &  $-1.0^{+0.0}_{-0.1}$ &  $-0.0^{+0.0}_{-0.1}$ &  $-0.4^{+0.3}_{-0.4}$ \\ 
090510 & $8.9^{+0.9}_{-2.2}$ &  $9.2^{+2.0}_{-1.5}$ &  $0.016^{+0.019}_{-0.010}$ &  $2.15^{+0.10}_{-0.07}$ &  $0.11^{+0.02}_{-0.02}$ &  $164^{+55}_{-60}$ &  $-2.0^{+0.3}_{-0.2}$ &  $-1.1^{+0.1}_{-0.2}$ &  $-0.1^{+0.1}_{-0.2}$ &  $-1.3^{+0.3}_{-0.2}$ \\ 
091109B & $7.7^{+1.6}_{-2.4}$ &  $9.1^{+2.0}_{-2.3}$ &  $0.049^{+0.028}_{-0.021}$ &  $2.13^{+0.10}_{-0.08}$ &  $0.11^{+0.02}_{-0.04}$ &  $140^{+65}_{-69}$ &  $1.6^{+0.9}_{-1.1}$ &  $-3.0^{+0.9}_{-1.1}$ &  $-1.4^{+0.9}_{-1.1}$ &  $-2.9^{+1.6}_{-1.9}$ \\ 
110112A & $7.1^{+2.1}_{-2.0}$ &  $8.6^{+2.1}_{-1.9}$ &  $0.044^{+0.029}_{-0.024}$ &  $2.13^{+0.13}_{-0.09}$ &  $0.12^{+0.02}_{-0.03}$ &  $122^{+74}_{-54}$ &  $-1.7^{+0.5}_{-0.3}$ &  $-1.3^{+0.2}_{-0.5}$ &  $-0.3^{+0.2}_{-0.5}$ &  $-1.0^{+0.4}_{-0.4}$ \\ 
111020A & $9.0^{+0.7}_{-1.6}$ &  $9.7^{+1.4}_{-1.4}$ &  $0.027^{+0.026}_{-0.016}$ &  $2.04^{+0.15}_{-0.03}$ &  $0.10^{+0.01}_{-0.02}$ &  $175^{+40}_{-51}$ &  $0.4^{+0.7}_{-0.8}$ &  $-1.6^{+0.4}_{-0.7}$ &  $-1.4^{+0.8}_{-0.8}$ &  $-0.7^{+0.5}_{-0.7}$ \\ 
111121A & $6.9^{+2.2}_{-2.8}$ &  $10.0^{+1.3}_{-2.0}$ &  $0.028^{+0.018}_{-0.013}$ &  $2.27^{+0.32}_{-0.19}$ &  $0.10^{+0.01}_{-0.02}$ &  $138^{+67}_{-72}$ &  $2.2^{+0.6}_{-0.9}$ &  $-3.2^{+0.6}_{-1.0}$ &  $-0.7^{+0.5}_{-0.8}$ &  $-2.9^{+1.3}_{-1.4}$ \\ 
121226A & $3.2^{+2.0}_{-0.8}$ &  $10.8^{+0.9}_{-1.4}$ &  $0.026^{+0.011}_{-0.011}$ &  $2.27^{+0.33}_{-0.19}$ &  $0.09^{+0.01}_{-0.01}$ &  $68^{+52}_{-24}$ &  $2.0^{+0.7}_{-1.4}$ &  $-2.2^{+0.7}_{-0.9}$ &  $-0.6^{+0.4}_{-0.8}$ &  $-3.2^{+1.8}_{-1.5}$ \\ 
% Mean &   7.8 &     8.6 &        0.057 &  2.26 &      0.12 &  135 & -0.3 & -1.7 &  -0.5 &  -1.7 \\
% Std  &   1.7 &     2.1 &        0.052 &  0.22 &      0.04 &   47 &  1.4 &  0.7 &   0.4 &   1.0 \\
Mean$\pm$Std\footnote{Means and standard deviations are calculated from the 14 cosmological short GRBs.} &   7.8$\pm$1.7 &     8.6$\pm$2.1 &  0.057$\pm$0.052 &  2.26$\pm$0.22 &      0.12$\pm$0.04 &  135$\pm$47 & $-0.3\pm$0.4 & $-1.7\pm$0.7 &  $-0.5\pm$0.4 &  $-1.7\pm$1.0 \\
\bottomrule
\end{tabular}
\end{table*}

 %The detailed MCMC analysis results, including marginalized posterior distributions and best-fitting light curves, can be found in the Appendix.

% Outflow structures are mainly related to $\gamma_B$, considering jet opening angles are approximated by $\theta_0 \sim 1/\gamma_B$. On-axis Lorentz factors of outflows are given by $\Gamma \sim 2\eta_0 \gamma_B$.

%As for the degenerate parameters, most of them are not well constrained and have broad posterior distributions due to scaling relations. We note that the uncertainties of degenerate parameters can be marginalized out, and thus, will not affect the constraints of non-degenerate parameters. In this work, we focus our analysis on ($\eta_0, \gamma_B, \theta_{\rm obs}, p$) and ($\theta_0, \Gamma$). 

%In terms of outflow structures, GW170817 and short GRBs share the same kind of outflows. 
For GW170817, the observation angle $\theta_{\rm obs} = 0.53$ radians $= 30^{\circ}$ is significantly larger than $\theta_0 = 0.11$ radians $= 6.3^{\circ}$, which indicates that it is viewed significantly off-axis, outside of the jet opening angle. For short GRBs, the mean and standard deviation of observation angles is 0.06 $\pm$ 0.05 radians ($3.4^{\circ} \pm 2.9^{\circ}$). In Figure \ref{fig:contour_theta}, short GRBs (blue squares) are located below the dash grey line ($\theta_{\rm obs} = \theta_0$), which indicates the line of sight is located inside the cone of the outflow. We note that GW170817 is the nearest event ($z = 0.00973$). After the detection of the gravitational wave signal, it attracted significant attention from the community and was monitored intensively. On the other hand, the 14 short GRBs are cosmological ($z = 0.12$-$2.6$). The significant difference of observation angles between GW170817 and cosmological short GRBs can be explained by observation bias. Most cosmological short GRBs are detected on-axis, otherwise, they would be too weak to be conclusively detected.

\begin{figure}[hbt] 
\centering
\includegraphics[width=\linewidth]{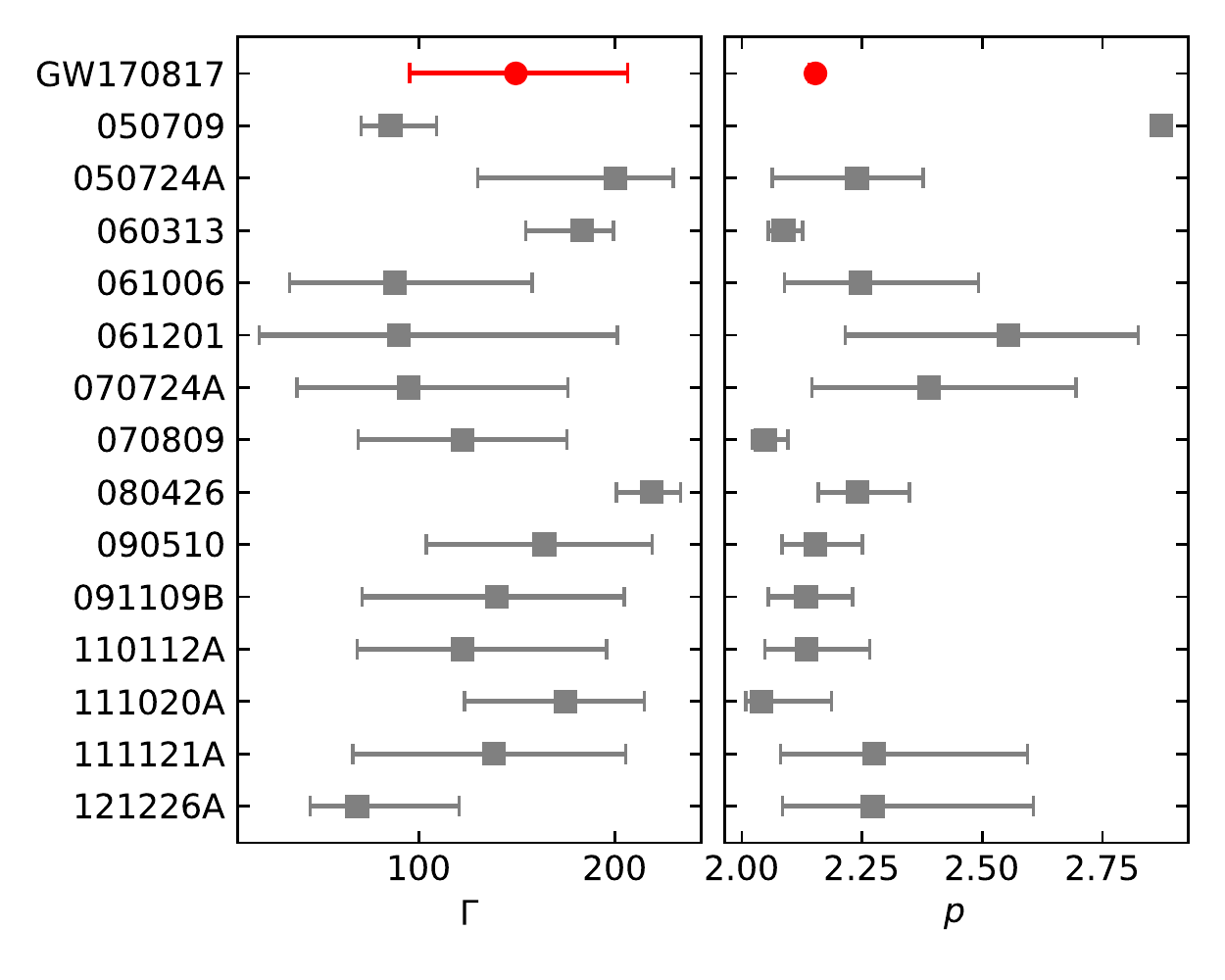}
\caption{Fit results of the characteristic Lorentz factor $\Gamma$ (left) and the spectral index $p$ (right) for GW170817 (red circle) and 14 short GRBs (blue squares). Markers with error bars indicate median values and symmetric 68\% quantiles. }
\label{fig:contour_gamma_p}
\end{figure}

In Figure \ref{fig:contour_gamma_p}, we show the fitting results of the characteristic Lorentz factor ($\Gamma \sim 2\eta_0\gamma_B$) for GW170817 and 14 short GRBs.  GW170817 has $\Gamma \approx 150$, which fits well within the range of short GRBs ($\Gamma = 135 \pm 47$).

% \begin{figure}[hbt] 
% \centering
% \includegraphics[width=\linewidth]{R_08_p.pdf}
% \caption{Fit results of spectral index $p$ for 14 short GRBs (blue squares) and GW170817 (red circle). Markers with error bars indicate median values and symmetric 68\% quantiles. }
% \label{fig:contour_p}
% \end{figure}

In Figure \ref{fig:contour_gamma_p}, we show the fitting results of the spectral index $p$. GW170817 has a tight constraint $p = 2.15^{+0.01}_{-0.01}$, since it has good observational data from all three bands. The mean and standard deviation of short GRBs is $p = 2.26 \pm 0.22$. The spectral index of GW170817 is consistent with those of short GRBs.

\begin{figure}[hbt] 
\centering
\includegraphics[width=\linewidth]{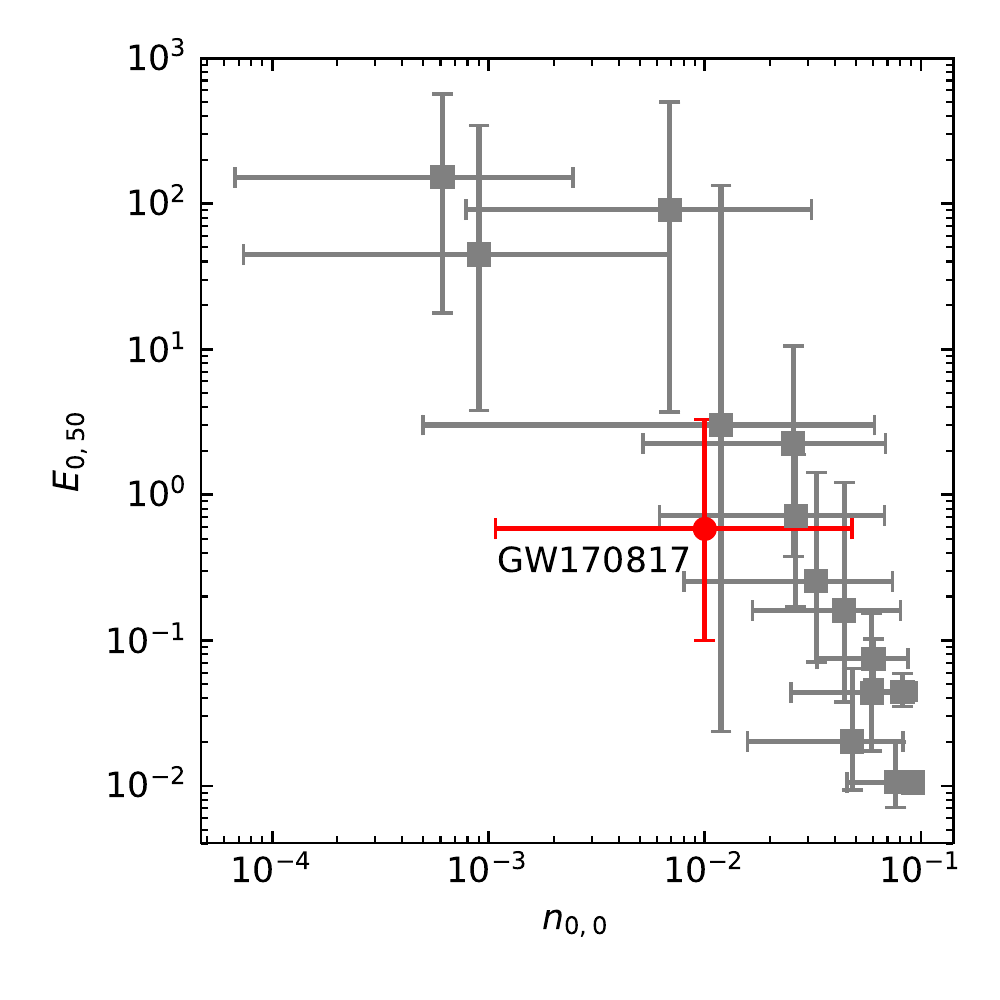}
\caption{Fit results of explosion energy $E_{0,50}$ and ISM density $n_{0,0}$ for 14 short GRBs (blue squares) and GW170817 (red circle). Markers with error bars indicate median values and symmetric 68\% quantiles. }
\label{fig:contour_nE}
\end{figure}

% \begin{figure}[hbt] 
% \centering
% \includegraphics[width=\linewidth]{R_08_eB.pdf}
% \caption{Fit results of $\epsilon_{e}$ and $\epsilon_{B}$ for 14 short GRBs (blue squares) and GW170817 (red circle). Markers with error bars indicate median values and symmetric 68\% quantiles. }
% \label{fig:contour_eB}
% \end{figure}

Figures \ref{fig:contour_nE} shows the distributions of GW170817 and 14 short GRBs in the ($n_{0,0}, E_{0,50}$) plane. GW170817 has $n_0 \approx 10^{-2}$ proton cm$^{-3}$ and $E_0 \approx 6 \times 10^{49}$ erg. Considering the jet opening angle $\theta_0 \approx 0.11$ radians, the corresponding isotropic equivalent energy is $E_{\rm iso} \approx 8 \times 10^{52}$ erg.  
These values are located within the typical ranges of short GRBs. Since $n_{0,0}$ and $E_{0,50}$ are degenerate parameters due to scaling relations, most of the bursts display large error bars. Even though these degenerate parameters  are not well constrained, we note that their uncertainties can be marginalized out, and thus, will not affect the constraints on non-degenerate parameters.

\subsection{On-axis Light Curves}
We have found that GW170817 and short GRBs share similar outflow structures and other physical parameters, except for the observation angle. Viewed off-axis, GW170817 displayed exceptional behaviors, such as late onset and early brightening. Short GRBs are viewed on-axis and show a general decline shortly after prompt emission. This leads to an interesting question: what would be observed from GW170817 if the observer were located on-axis? 

\begin{figure}[hbt] 
\centering
\includegraphics[width=\linewidth]{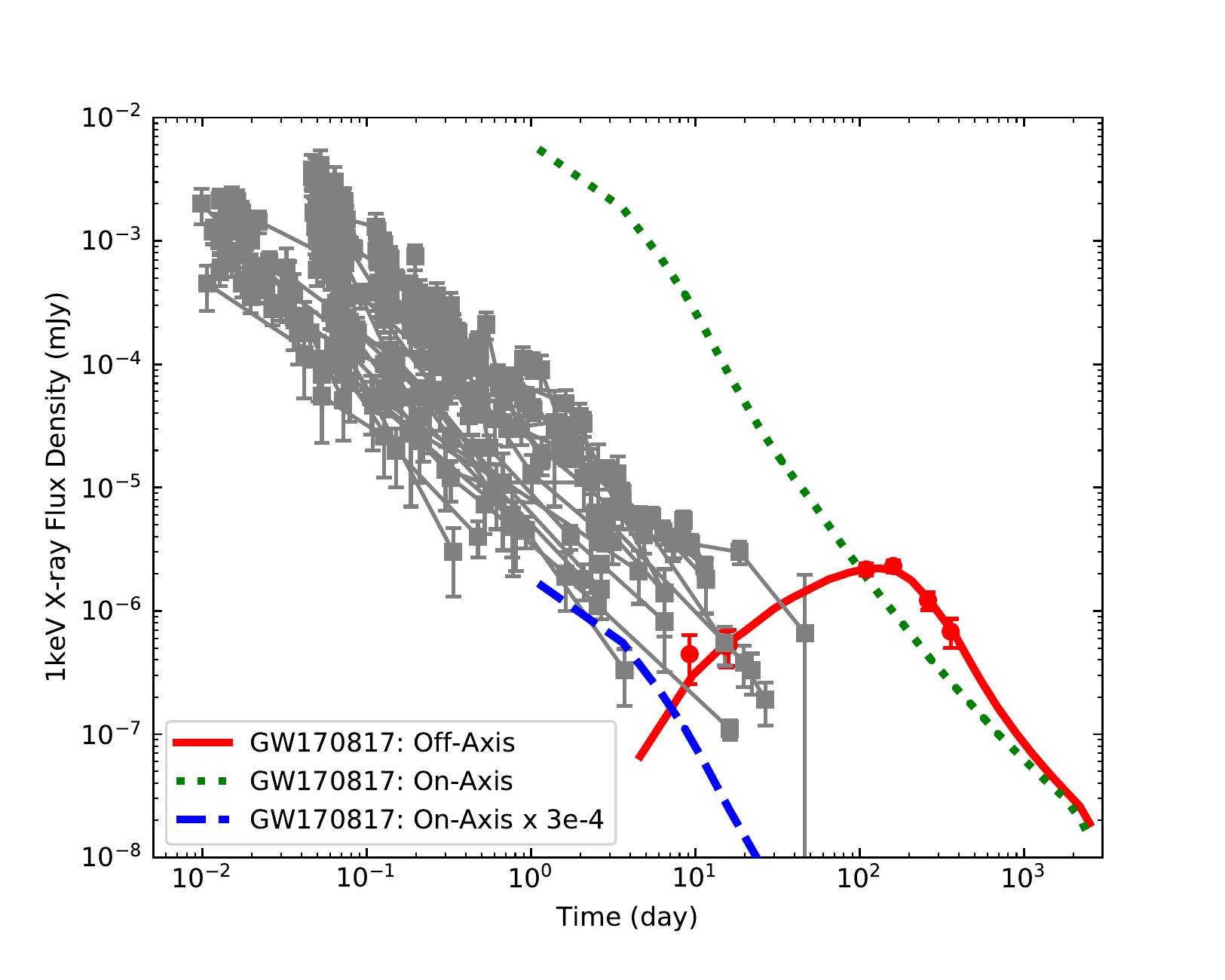}
\caption{X-ray afterglow observations for GW170817 (red circles with error bars) and 27 cosmological short GRBs (grey squares with error bars). Red solid line: the best-fitting light curve ($\theta_{\rm obs} = 0.53$ radians $= 30^{\circ}$). Green dotted line: the on-axis light curve obtained by setting $\theta_{\rm obs} = 0$. Blue dashed line: the on-axis light curve adjusted by an inverse square distance factor of $3\times10^{-4}$. }
\label{fig: lc on-axis}
\end{figure}

In Figure \ref{fig: lc on-axis}, we show the X-ray afterglow observations for GW170817 (red circles with error bars) and 27 cosmological short GRBs (grey squares with error bars). The best-fitting light curve for GW170817 (red solid line) fits the observational data very well. It captures the late onset at around $\sim 9$ days, the steady brightening up to $\sim 100$ days and the turnover at $\sim 150$ days. 

Given the set of best-fitting parameters for GW170817, we can generate the on-axis light curve by setting $\theta_{\rm obs} = 0$ and leaving all other parameters unchanged. The resulting on-axis light curve is shown in green dotted line in Figure \ref{fig: lc on-axis}. It shows a monotonic decline, just like other short GRBs. At  late times, the on-axis light curve coincides with the off-axis light curve. This is due to the whole region of the decelerated outflow becoming observable for both on-axis and off-axis observers. 

Since GW170817 is a local event ($d_L = 39.5$ Mpc), its flux density is significantly higher than others. The median redshift for short GRBs is $z = 0.46$ \citep{fong2017electromagnetic}.  Using a benchmark $\Lambda CDM$ cosmology with $H_0 = 71$ km s$^{-1}$Mpc$^{-1}$ and $\Omega_m = 0.27$,  the corresponding luminosity distance can be calculated as $d_L \sim 2500$ Mpc. The  inverse square factor can be roughly estimated as $3 \times 10^{-4}$.  The on-axis light curve adjusted for the inverse square factor is shown as the blue dashed line. Though located a little lower, it is consistent with the observations from short GRBs. This reveals that GW170817 is similar to short GRBs. 

\section{Discussion} \label{sec: GRB discussion}
We systematically compare the properties of GW170817 and a population of short GRBs by performing MCMC analysis in the 8-dimension parameter space of hydrodynamic, radiation and observational parameters,  

% We note that, in \citet{wu2018constraining}, we mistook the peaks of posterior distributions as medians.  In this letter, we correct it as 50\% quantiles. The values of fitting parameters could vary slightly depending on the shape of distributions. The main conclusion for GW170817, a relativistic structured jet viewed off-axis, remains unaffected.

We demonstrate that GW170817 and short GRBs share the same outflow structure: a relativistic structured jet. The only difference in our analysis between GW170817 and the cosmological short GRBs is that GW1701817 is viewed off-axis and the cosmological short GRBs are viewed on-axis. The difference in observation angle can explain the exceptional behavior of the GW170817 afterglow light curve, such as the late onset and early brightening.  Other properties of the GW170817 afterglow, including jet opening angle, Lorentz factor and spectral index, are all consistent with those of cosmological short GRBs. 

We calculate the light curve for the GW170817 afterglow that on-axis viewers would have observed. It shows a temporal decline consistent with cosmological short GRBs. The similarity between GW170817 and short GRBs indicates that cosmological short GRBs are also neutron star mergers.

%Thus, GW170817 is not a new transient of low luminosity, but a regular short GRB viewed off-axis. 
%not intrinsically subluminous and its exceptional properties result from a different viewing angle than classical short GRBs. The similarity  between  GW170817  the  and  short  GRB  population  indicates  that  short  GRBs originate from binary neutron star mergers and have jet-like outflow structures. 

\section{Acknowledgements}
We are grateful to Michael Blanton, Wen-fai Fong and Roman Scoccimarro for helpful discussions and comments. 

\bibliography{bibliography.bib}

\begin{thebibliography}{}
\expandafter\ifx\csname natexlab\endcsname\relax\def\natexlab#1{#1}\fi
\providecommand{\url}[1]{\href{#1}{#1}}

\bibitem[{Abbott {et~al.}(2017)Abbott, Abbott, Abbott, Acernese, Ackley, Adams,
  Adams, Addesso, Adhikari, Adya, {et~al.}}]{Abbott2017}
Abbott, B.~P., Abbott, R., Abbott, T., {et~al.} 2017, Physical Review Letters,
  119, 161101

\bibitem[{Alexander {et~al.}(2017)Alexander, Berger, Fong, Williams, Guidorzi,
  Margutti, Metzger, Annis, Blanchard, Brout, {et~al.}}]{Alexander2017}
Alexander, K., Berger, E., Fong, W., {et~al.} 2017, The Astrophysical Journal
  Letters, 848, L21

\bibitem[{Alexander {et~al.}(2018)Alexander, Margutti, Blanchard, Fong, Berger,
  Hajela, Eftekhari, Chornock, Cowperthwaite, Giannios, Guidorzi,
  Kathirgamaraju, MacFadyen, Metzger, Nicholl, Sironi, Villar, Williams, Xie,
  \& Zrake}]{Alexander2018}
Alexander, K.~D., Margutti, R., Blanchard, P.~K., {et~al.} 2018, The
  Astrophysical Journal Letters, 863, L18

\bibitem[{Beniamini {et~al.}(2018)Beniamini, Petropoulou, Duran, \&
  Giannios}]{beniamini2018lesson}
Beniamini, P., Petropoulou, M., Duran, R.~B., \& Giannios, D. 2018, arXiv
  preprint arXiv:1808.04831

\bibitem[{Bromberg {et~al.}(2017)Bromberg, Tchekhovskoy, Gottlieb, Nakar, \&
  Piran}]{Bromberg2017}
Bromberg, O., Tchekhovskoy, A., Gottlieb, O., Nakar, E., \& Piran, T. 2017,
  Monthly Notices of the Royal Astronomical Society, 475, 2971

\bibitem[{{D\'{}Avanzo, P.} {et~al.}(2018){D\'{}Avanzo, P.}, {Campana, S.},
  {Salafia, O. S.}, {Ghirlanda, G.}, {Ghisellini, G.}, {Melandri, A.},
  {Bernardini, M. G.}, {Branchesi, M.}, {Chassande-Mottin, E.}, {Covino, S.},
  {D\'{}Elia, V.}, {Nava, L.}, {Salvaterra, R.}, {Tagliaferri, G.}, \&
  {Vergani, S. D.}}]{d2018evidence}
{D\'{}Avanzo, P.}, {Campana, S.}, {Salafia, O. S.}, {et~al.} 2018, A\&A, 613,
  L1

\bibitem[{Dobie {et~al.}(2018)Dobie, Kaplan, Murphy, Lenc, Mooley, Lynch,
  Corsi, Frail, Kasliwal, \& Hallinan}]{Dobie2018}
Dobie, D., Kaplan, D.~L., Murphy, T., {et~al.} 2018, The Astrophysical Journal
  Letters, 858, L15

\bibitem[{Duffell \& MacFadyen(2013{\natexlab{a}})}]{Duffell2013RT}
Duffell, P.~C., \& MacFadyen, A.~I. 2013{\natexlab{a}}, The Astrophysical
  Journal, 775, 87

\bibitem[{Duffell \& MacFadyen(2013{\natexlab{b}})}]{Duffell2013}
---. 2013{\natexlab{b}}, The Astrophysical Journal Letters, 776, L9

\bibitem[{Fong {et~al.}(2015)Fong, Berger, Margutti, \&
  Zauderer}]{fong2015decade}
Fong, W.-f., Berger, E., Margutti, R., \& Zauderer, B.~A. 2015, The
  Astrophysical Journal, 815, 102

\bibitem[{Fong {et~al.}(2017)Fong, Berger, Blanchard, Margutti, Cowperthwaite,
  Chornock, Alexander, Metzger, Villar, Nicholl,
  {et~al.}}]{fong2017electromagnetic}
Fong, W.-f., Berger, E., Blanchard, P., {et~al.} 2017, The Astrophysical
  Journal Letters, 848, L23

\bibitem[{Foreman-Mackey {et~al.}(2013)Foreman-Mackey, Hogg, Lang, \&
  Goodman}]{Foreman2013}
Foreman-Mackey, D., Hogg, D.~W., Lang, D., \& Goodman, J. 2013, Publications of
  the Astronomical Society of the Pacific, 125, 306

\bibitem[{Gill \& Granot(2018)}]{Gill2018}
Gill, R., \& Granot, J. 2018, Monthly Notices of the Royal Astronomical Society

\bibitem[{Goldstein {et~al.}(2017)Goldstein, Veres, Burns, Briggs, Hamburg,
  Kocevski, Wilson-Hodge, Preece, Poolakkil, Roberts, {et~al.}}]{Goldstein2017}
Goldstein, A., Veres, P., Burns, E., {et~al.} 2017, The Astrophysical Journal
  Letters, 848, L14

\bibitem[{Goodman {et~al.}(2010)Goodman, Weare, {et~al.}}]{Goodman2010}
Goodman, J., Weare, J., {et~al.} 2010, Communications in applied mathematics
  and computational science, 5, 65

\bibitem[{Gottlieb {et~al.}(2017)Gottlieb, Nakar, \& Piran}]{Gottlieb2018}
Gottlieb, O., Nakar, E., \& Piran, T. 2017, Monthly Notices of the Royal
  Astronomical Society, 473, 576

\bibitem[{Haggard {et~al.}(2017)Haggard, Nynka, Ruan, Kalogera, Cenko, Evans,
  \& Kennea}]{Haggard2017}
Haggard, D., Nynka, M., Ruan, J.~J., {et~al.} 2017, The Astrophysical Journal
  Letters, 848, L25

\bibitem[{Hallinan {et~al.}(2017)Hallinan, Corsi, Mooley, Hotokezaka, Nakar,
  Kasliwal, Kaplan, Frail, Myers, Murphy, {et~al.}}]{Hallinan2017}
Hallinan, G., Corsi, A., Mooley, K., {et~al.} 2017, Science, eaap9855

\bibitem[{Hotokezaka {et~al.}(2018)Hotokezaka, Kiuchi, Shibata, Nakar, \&
  Piran}]{Hotokezaka2018}
Hotokezaka, K., Kiuchi, K., Shibata, M., Nakar, E., \& Piran, T. 2018, arXiv
  preprint arXiv:1803.00599

\bibitem[{Kasliwal {et~al.}(2017)Kasliwal, Nakar, Singer, Kaplan, Cook,
  Van~Sistine, Lau, Fremling, Gottlieb, Jencson, {et~al.}}]{Kasliwal2017}
Kasliwal, M., Nakar, E., Singer, L., {et~al.} 2017, Science, 358, 1559

\bibitem[{Kathirgamaraju {et~al.}(2017)Kathirgamaraju, Barniol~Duran, \&
  Giannios}]{Kathirgamaraju2017}
Kathirgamaraju, A., Barniol~Duran, R., \& Giannios, D. 2017, Monthly Notices of
  the Royal Astronomical Society: Letters, 473, L121

\bibitem[{Lamb {et~al.}(2019)Lamb, Lyman, Levan, Tanvir, Kangas, Fruchter,
  Gompertz, Hjorth, Mandel, Oates, {et~al.}}]{lamb2019optical}
Lamb, G., Lyman, J., Levan, A., {et~al.} 2019, The Astrophysical Journal
  Letters, 870, L15

\bibitem[{Lamb \& Kobayashi(2017)}]{Lamb2017}
Lamb, G.~P., \& Kobayashi, S. 2017, Monthly Notices of the Royal Astronomical
  Society, 472, 4953

\bibitem[{Lamb {et~al.}(2018)Lamb, Mandel, \& Resmi}]{lamb2018late}
Lamb, G.~P., Mandel, I., \& Resmi, L. 2018, Monthly Notices of the Royal
  Astronomical Society, 481, 2581

\bibitem[{Lazzati {et~al.}(2018)Lazzati, Perna, Morsony, Lopez-Camara,
  Cantiello, Ciolfi, Giacomazzo, \& Workman}]{lazzati2017}
Lazzati, D., Perna, R., Morsony, B.~J., {et~al.} 2018, Physical Review Letters,
  120, 241103

\bibitem[{Lyman {et~al.}(2018)Lyman, Lamb, Levan, Mandel, Tanvir, Kobayashi,
  Gompertz, Hjorth, Fruchter, Kangas, Steeghs, Steele, Cano, Copperwheat,
  Evans, Fynbo, Gall, Im, Izzo, Jakobsson, Milvang-Jensen, O'Brien, Osborne,
  Palazzi, Perley, Pian, Rosswog, Rowlinson, Schulze, Stanway, Sutton,
  Th{\"o}ne, de~Ugarte~Postigo, Watson, Wiersema, \& Wijers}]{Lyman2018}
Lyman, J.~D., Lamb, G.~P., Levan, A.~J., {et~al.} 2018, Nature Astronomy, 2,
  751

\bibitem[{Margutti {et~al.}(2017)Margutti, Berger, Fong, Guidorzi, Alexander,
  Metzger, Blanchard, Cowperthwaite, Chornock, Eftekhari,
  {et~al.}}]{Margutti2017}
Margutti, R., Berger, E., Fong, W., {et~al.} 2017, The Astrophysical Journal
  Letters, 848, L20

\bibitem[{Margutti {et~al.}(2018)Margutti, Alexander, Xie, Sironi, Metzger,
  Kathirgamaraju, Fong, Blanchard, Berger, MacFadyen, {et~al.}}]{Margutti2018}
Margutti, R., Alexander, K., Xie, X., {et~al.} 2018, The Astrophysical Journal
  Letters, 856, L18

\bibitem[{Mooley {et~al.}(2018{\natexlab{a}})Mooley, Nakar, Hotokezaka,
  Hallinan, Corsi, Frail, Horesh, Murphy, Lenc, Kaplan,
  {et~al.}}]{mooley2018mildly}
Mooley, K., Nakar, E., Hotokezaka, K., {et~al.} 2018{\natexlab{a}}, Nature,
  554, 207

\bibitem[{Mooley {et~al.}(2018{\natexlab{b}})Mooley, Frail, Dobie, Lenc, Corsi,
  De, Nayana, Makhathini, Heywood, Murphy, {et~al.}}]{mooley2018strong}
Mooley, K., Frail, D., Dobie, D., {et~al.} 2018{\natexlab{b}}, The
  Astrophysical Journal Letters, 868, L11

\bibitem[{Nakar {et~al.}(2018)Nakar, Gottlieb, Piran, Kasliwal, \&
  Hallinan}]{nakar2018gamma}
Nakar, E., Gottlieb, O., Piran, T., Kasliwal, M.~M., \& Hallinan, G. 2018,
  arXiv preprint arXiv:1803.07595

\bibitem[{Nynka {et~al.}(2018)Nynka, Ruan, Haggard, \& Evans}]{nynka2018fading}
Nynka, M., Ruan, J.~J., Haggard, D., \& Evans, P.~A. 2018, The Astrophysical
  Journal Letters, 862, L19

\bibitem[{Piro {et~al.}(2018)Piro, Troja, Zhang, Ryan, Van~Eerten, Ricci,
  Wieringa, Tiengo, Butler, Cenko, {et~al.}}]{piro2018long}
Piro, L., Troja, E., Zhang, B., {et~al.} 2018, Monthly Notices of the Royal
  Astronomical Society, 483, 1912

\bibitem[{Resmi {et~al.}(2018)Resmi, Schulze, Ishwara-Chandra, Misra, Buchner,
  De~Pasquale, S{\'a}nchez-Ram{\'\i}rez, Klose, Kim, Tanvir,
  {et~al.}}]{Resmi2018}
Resmi, L., Schulze, S., Ishwara-Chandra, C., {et~al.} 2018, The Astrophysical
  Journal, 867, 57

\bibitem[{Ruan {et~al.}(2018)Ruan, Nynka, Haggard, Kalogera, \&
  Evans}]{ruan2018brightening}
Ruan, J.~J., Nynka, M., Haggard, D., Kalogera, V., \& Evans, P. 2018, The
  Astrophysical Journal Letters, 853, L4

\bibitem[{Ryan {et~al.}(2015)Ryan, Van~Eerten, MacFadyen, \& Zhang}]{Ryan2015}
Ryan, G., Van~Eerten, H., MacFadyen, A., \& Zhang, B.-B. 2015, The
  Astrophysical Journal, 799, 3

\bibitem[{Savchenko {et~al.}(2017)Savchenko, Ferrigno, Kuulkers, Bazzano,
  Bozzo, Brandt, Chenevez, Courvoisier, Diehl, Domingo,
  {et~al.}}]{Savchenko2017}
Savchenko, V., Ferrigno, C., Kuulkers, E., {et~al.} 2017, The Astrophysical
  Journal Letters, 848, L15

\bibitem[{Troja {et~al.}(2017)Troja, Piro, van Eerten, Wollaeger, Im, Fox,
  Butler, Cenko, Sakamoto, Fryer, {et~al.}}]{Troja2017}
Troja, E., Piro, L., van Eerten, H., {et~al.} 2017, Nature, 551, 71

\bibitem[{Troja {et~al.}(2018)Troja, Piro, Ryan, van Eerten, Ricci, Wieringa,
  Lotti, Sakamoto, \& Cenko}]{Troja2018}
Troja, E., Piro, L., Ryan, G., {et~al.} 2018, Monthly Notices of the Royal
  Astronomical Society: Letters, 478, L18

\bibitem[{van Eerten {et~al.}(2018)van Eerten, Ryan, Ricci, Burgess, Wieringa,
  Piro, Cenko, \& Sakamoto}]{van2018year}
van Eerten, E. T.~H., Ryan, G., Ricci, R., {et~al.} 2018, arXiv preprint
  arXiv:1808.06617

\bibitem[{Van~Eerten \& MacFadyen(2012)}]{Eertenn2012Scale}
Van~Eerten, H.~J., \& MacFadyen, A.~I. 2012, The Astrophysical Journal Letters,
  747, L30

\bibitem[{Wu \& MacFadyen(2018)}]{wu2018constraining}
Wu, Y., \& MacFadyen, A. 2018, The Astrophysical Journal, 869, 55

\bibitem[{Xie {et~al.}(2018)Xie, Zrake, \& MacFadyen}]{Xie2018}
Xie, X., Zrake, J., \& MacFadyen, A. 2018, The Astrophysical Journal, 863, 58

\bibitem[{{Zrake} {et~al.}(2018){Zrake}, {Xie}, \&
  {MacFadyen}}]{2018ApJ...865L...2Z}
{Zrake}, J., {Xie}, X., \& {MacFadyen}, A. 2018, \apjl, 865, L2

\end{thebibliography}

\end{document}